\documentstyle[proceedings]{jd} 

\def\spose#1{\hbox to 0pt{#1\hss}}
\def\lta{\mathrel{\spose{\lower 3pt\hbox{$\mathchar"218$}}
     \raise 2.0pt\hbox{$\mathchar"13C$}}}
\def\gta{\mathrel{\spose{\lower 3pt\hbox{$\mathchar"218$}}
     \raise 2.0pt\hbox{$\mathchar"13E$}}}
\def\etal{{\it et al.\ }}

\begin{opening}
\title{ABUNDANCE RATIOS IN COMPOSITE STELLAR POPULATIONS \protect\\
       with special emphasis on 
       ELLIPTICAL GALAXIES}

\author{Uta Fritze - von Alvensleben}
\institute{Universit\"atssternwarte G\"ottingen\\
           Geismarlandstr. 11, 37083 G\"ottingen, Germany}
\end{opening}

\runningtitle{ABUNDANCE RATIOS IN COMPOSITE STELLAR POPULATIONS}

\begin{document}

\section{Introduction}

Abundance Ratios are observed in  a wide variety of instances 
addressed in this volume: in {\bf Gas} and {\bf Stars}, 
Simple Stellar Populations ({\bf SSPs}) like star clusters, and, finally, in 
Composite Stellar Populations ({\bf CSPs}) like E/S0 galaxies (Bulges, spheroidal galaxies, 
Merger Remnants, etc.) on which the focus of this contribution will be. I'll 
briefly review the observational situation and the current interpretation and then cast 
some doubt on the latter from our own chemically consistent spectrophotometric 
modelling results. 
 
\section{Observations and Standard Interpretation}

As defined in the Lick system, absorption line indices like Fe5270, Fe5335, 
Mg$_2$, Mgb, 
and many others have been measured by various groups on large numbers of early type 
galaxies and bulges. 
For the nuclei of giant field ellipticals Fe5270 and Fe5335 are seen to correlate 
with Mg$_2$, but the slopes of Fe5270 and Fe5335 versus 
Mg$_2$ are found to be rather shallow with considerable scatter, they do 
not follow the relation for globular clusters, and they are flatter 
for nuclei of different galaxies than within one galaxy (see e.g. Faber \etal 1992, 
Worthey \etal 1992, Davies \etal 1993). {\bf Gradients} in Mg$_2$ tend to be 
stronger for larger Es, gradients in Fe correlate with gradients in Mg$_2$. 
Mg$_2$ correlates with velocity dispersion $\sigma$ while 
$\langle {\rm Fe}\rangle := \frac{1}{2} ({\rm Fe5270 + Fe5335})$ does not 
(or only weakly). Ellipticals with fine 
structure or peculiar core kinematics tend to have lower Mg$_2$ than 
predicted for their luminosity by the luminosity -- metallicity relation  
(see e.g. Schweizer \etal 1990, Bender \etal 1993, Carollo \& Danziger 
1994).

For a large sample of E/S0s in nearby {\bf clusters} J{\o}rgensen 1997 confirms 
the trends for field ellipticals and shows that 
Mg$_2$ and $\langle {\rm Fe}\rangle$ both correlate with projected 
cluster surface density and that 
Mg$_2$ strongly correlates with M/L while 
$\langle {\rm Fe}\rangle$ does so only weakly.

Comparing the relative strengths of Fe5270, Fe5335, and Mg indices 
%usually expressed in terms of  ${\rm [MgFe]_{\ast} ~ := ~ Log~ \sqrt{Mgb \cdot 
%\langle Fe\rangle}}$, 
measured on CSPs with those observed in Galactic 
reference stars generally leads to the conclusion that {\bf magnesium seems to 
be enhanced with respect to iron} in several cases. 
Typical values quoted for the nuclei of massive boxy {\bf Ellipticals} are 
[Mg/Fe] $\sim 0.4 - 0.6$, while 
[Mg/Fe] $\sim 0$ for the lower luminosity disky ellipticals. 
[Mg/Fe] seems to increase with increasing velocity 
dispersion, luminosity, metallicity, and -- for cluster ellipticals -- with 
increasing cluster surface density, while inside galaxies [Mg/Fe] seems to 
be constant (see Bender 1997 for an extensive review). 
For {\bf Bulges} the situation is less clear. While Sadler \etal 1996 find 
[Mg/Fe] $\sim 0$ \ for [Fe/H] $\geq 0$ and [Mg/Fe] $\sim +0.2$ for [Fe/H] $< 0$, 
Jablonka \etal 1996 report a range of [Mg/Fe] values from $0$ to $+0.5$ with no 
difference between 
low and high luminosity bulges. {\bf Spheroidal Galaxies} have 
[Mg/Fe] $\sim 0$ (Gorgas \etal 1997), the centers of luminous {\bf S0s} resemble 
ellipticals with [Mg/Fe] $> 0$ (Fisher \etal 1996).

\noindent
The {\bf Standard Interpretation} of these results relies on the assumptions 
that 
\begin{enumerate}
\item relative strengths 
of absorption indices reflect element abundance ratios in the stellar 
population and 
\item that these stellar element ratios reflect abundance ratios in the 
ISM out of which the stars were formed. 
\end{enumerate}

If, and only if, both assumptions were correct (I'll show below that this is not the case for 
CSPs), an observed stellar [Mg/Fe]$_{\ast} > 0$ 
would 
reflect an ISM abundance ratio [Mg/Fe]$_{\rm ISM} > 0$ 
at the birth of (the bulk of) the stellar population. 

\noindent
Mg, like other 
$\alpha$-elements, is a typical SNII product 
that is returned to the ISM with the death of the first massive stars 
on a timescale of $10^6$ yr. Fe has significant ($\gta 50 \%$) contributions 
from SNIa, it is restored to the ISM on a timescale of 
several $10^8$ yr (see e.g. Matteucci 1994). 

\noindent
[Mg/Fe]$_{\rm ISM} > 0$ would have either (or several) of the following implications 
for the {\bf central regions} of galaxies: 
\begin{itemize}
\item an IMF favoring High Mass Star Formation ({\bf SF}), at least in early evolutionary 
stages or for a stellar Pop3 
\item a characteristic timescale for SF \ t$_{\ast}~< ~{\rm t_{SNI}} \sim 3 \cdot 10^8$ yr 
\item a lower fraction of SNI progenitor binary stars
\item mass segregation and incomplete mixing (since only nuclear regions are observed)
\item low metallicity stellar yields (favour Mg over Fe for SNII)
\end{itemize}
Whatever effect is invoked, it must be stronger in massive Es, S0s (and possibly 
bulges) than in low mass systems (see e.g. Faber \etal 1992, Worthey \etal 1992, Matteucci 1994, 
Bender \etal 1993, 
Thomas \etal 1997 {\sl in prep}). The SF timescale t$_{\ast} \sim 3 \cdot 10^8$ yr is much shorter 
than the collapse time of order $\sim 1$ Gyr in the classical initial collapse SF scenario for ellipticals 
(Larson 1974), it reminds of the duration of starbursts in spiral-spiral mergers or 
of the lifetime of the Infrared Ultraluminous phase. 
Intriguingly, ASCA measurements seem to imply [Mg/Fe] $>0$ for the hot ICM in 
galaxy clusters despite the fact that the timescale for ICM enrichment is of the 
order of a Hubble time.

Stellar abundance ratios in CSPs depend on the SF history, the IMF, any eventual 
pre-enrichment (e.g. Pop3, halo $\rightarrow$ disk) or dilution (metal-poor 
infall) of the ISM 
{\bf before} the birth of the stars dominating the light at the index 
wavelengths today {\bf 
and} they also depend on possible modifications inside the stars 
through nucleosynthesis and mixing. 
Clearly, abundance ratios of the sun or solar neighborhood reference stars reflect the local 
SF history, IMF, and all possible pre-enrichment and dilution effects of the local ISM. 
A serious complication comes from the fact that e.g. the iron index 
$\langle {\rm Fe} \rangle$ 
is as sensitive to iron as it is to global metallicity (which is dominated by the 
SNII product oxygen), as shown by Tripicco \& Bell 1995.

\section{Chemically Consistent Modelling of Galaxy Evolution}

Starting from a gas cloud with primordial abundances and specifying a SF rate 
(as a function of time or current gas content) 
and an IMF -- the two fundamental 
parameters of our model -- successive generations of stars are formed with 
metallicities increasing from Z $=0$ to some present value. 
Their distribution over the HRD is calculated from stellar evolutionary tracks for 
various metallicities.  
Colour and absorption index calibrations in terms of (T$_{\rm eff}$, log g, 
[Fe/H]) and a library of model atmosphere spectra for various metallicities are 
used to describe the photometric and spectral evolution in a chemically 
consistent way (Einsel \etal 1995, M\"oller \etal 1997). 
Stellar yields are taken from nucleosynthesis models for a variety of 
metallicities. Solving a 
modified form of Tinsley's equations including SNI contributions from carbon 
deflagration white dwarf binaries in the way prescribed by Matteucci 1994 we 
obtain the time evolution of ISM abundances and abundance ratios for a series of 
elements. We properly account for the finite lifetime of each star, i.e. we do not 
use an Instantaneous Recycling Approximation. 
The SF history appropriate for the type of galaxy together with the IMF determine the evolution 
of ISM abundances and abundance ratios, the age and metallicity distributions, 
and the colour and index evolution of a composite stellar population. 

While by now the first reasonably complete sets of stellar evolutionary tracks, 
yields and model atmospheres for various metallicities are becoming available, 
the influence of non-solar abundance ratios is still far from being explored. 
A few recent efforts include Barbuy 1994 and Barbuy \etal 1995 who take the relative 
proportions of 
stars in various evolutionary stages from observed CMDs of globular clusters 
with various [Mg/Fe] and combine with Kurucz's model atmospheres 
(only available for [Mg/Fe] $= 0$). Borges \etal 1995 give calibrations for Mg$_2$ and NaD 
in terms of (T$_{\rm eff}$, log g, [Fe/H], {\bf and [Mg/Fe]}), and Weiss \etal 1995 show 
that for a 0.9 M$_{\odot}$ star the transition from an initial composition 
with [Mg/Fe] $=0$ to [Mg/Fe] 
$= + 0.4$ increases its Mg$_2$ index at fixed metallicity by as little as 0.03. 

No complete grid in 
stellar mass, [Mg/Fe], or [$\alpha$/Fe] is available, neither for stellar 
evolutionary tracks, nor for yields, colour or index calibrations, not to mention 
model atmospheres. 

A problem with stellar yields for non-solar metallicities is that they 
depend on several poorly known quantities, as e.g. $\Delta Y / \Delta Z$, 
explosion energies, mass loss rates, remnant masses, etc., and so, 
might not be very reliable yet. Beyond the discrepancies between various authors there 
is agreement, however, that for type II SNe the ratio of the Mg and the Fe yields, 
integrated over any plausible IMF, increases considerably with decreasing metallicity. 
Chemically consistent modelling of the ISM abundance evolution thus results in higher 
${\rm [Mg/Fe]_{ISM}}$ ratios in early evolutionary phases as compared to models 
using solar metallicity yields only. If the bulk of stars in an elliptical were 
formed on a short timescale we would expect them to have an enhanced Mg-to-Fe ratio 
frozen in. If this were then reflected in the stellar absorption line ratio of its 
CSP is an open question (see {\it caveat}s below and also Tantalo \etal 1997). 
In a spiral-spiral 
merger scenario for the formation of an E/S0 galaxy including a strong burst 
of SF the low metallicity stars of the gas-rich late-type spiral progenitors 
are all expected to have incorporated ${\rm [Mg/Fe]_{ISM} > 0}$ and the stars 
forming in the burst on a timescale of $\sim 10^8$ yr will have built in the 
rapidly increasing ${\rm [Mg/Fe]_{ISM}}$ during the burst (Fritze - v. A. 
\& Gerhard 1994a). This effect is seen in the 
spectra of young star clusters (SSPs) in the merger remnant NGC 7252 (Schweizer \& Seitzer 1993, 
Fritze - v. A. \& Burkert 1995). 

With appropriate SF histories our models for the evolution of ISM abundances and abundance ratios 
were used with solar metallicity yields to interprete the redshift evolution of 
MgII- and CIV- QSO absorption line systems that originate in galaxy haloes 
(Fritze - v. A. \etal 1989, 1991) and of Damped Ly$\alpha$ Absorbers (= DLAs) presumed to 
be (proto-)galactic disks (Fritze - v. A. \& Fricke 1995). First results from our chemically 
consistent models in comparison to DLA abundances are presented by Lindner \etal 1998.

\section{Results for Absorption Indices}
For SSPs of various metallicities we obtain the time evolution of colours, luminosities and 
indices. Using 
colours together with absorption indices allows 
to largely disentangle the age-metallicity degeneracy, 
since -- as first shown by Worthey 94 -- each colour 
and each index has its specific sensitivities to age and metallicity 
(Fritze -v. A. \& Burkert 1995). Model results include theoretical 
calibrations of indices, 
as e.g. Mg$_2$, in terms of [Fe/H] (Kurth \etal 1997, {\sl in prep.}).

For the CSP of an elliptical galaxy, our model e.g. for a ``classical L$^{\ast}$ -- 
Elliptical'' (1 zone, no dynamics with a characteristic timescale for SF t$_{\ast} = 1$ 
Gyr as indicated for an initial collapse 
scenario) gives a broad 
metallicity distribution with a luminosity-weighted mean metallicity in the 
V-band ${\rm \langle Z \rangle_V \sim \frac{1}{2} \cdot Z_{\odot}}$. While the metallicity distribution of our global galaxy model 
falls off sharply towards higher than solar metallicities it has a broad wing 
towards metallicities lower than ${\rm \frac{1}{2} \cdot Z_{\odot}}$
(M\"oller \etal 1997). Similar results from independent modelling approaches are reported by 
Greggio 1997 and Arimoto 1997 ({\sl this Joint Discussion}). 
The stellar metallicity you see depends on the wavelength of your
observation. For an elliptical, it is higher by $\sim 40 \%$ in K than it is in the V-band. Needless to say, 
luminosity weighted mean stellar metallicities in general differ from the ISM metallicity. 
Taking a consistent metallicity 
distribution into account, our model reaches the observed 
range of ${\rm [MgFe]_{\ast}  ~ := ~ Log~ \sqrt{Mgb \cdot 
\langle Fe \rangle} \sim 0.4 - 0.6}$ (see e.g. Gonzalez 1993) {\bf without} 
[Mg/Fe] $>$ 0, neither for stellar tracks, nor for index calibrations 
(M\"oller \etal 1997). With stellar evolutionary tracks from the Padova group, our chemically 
consistent ``classical elliptical galaxy'' model reaches ${\rm Mg_2 \geq 0.30}$ and 
${\rm Fe5335 \geq 3.1}$ at ages $\geq 15$ Gyr. Typical values observed in the nuclei of 
E/S0 galaxies are in the range ${\rm 0.18 \leq Mg_2 \leq 0.37}$ and 
${\rm 2.0 \leq Fe5335 \leq 3.5}$ (Worthey \etal 1992, Gonzalez 1993). 
In comparison with observational data it should 
be kept in mind that those always refer to central regions of galaxies 
(${\rm R < (\frac{1}{2} - 1) \cdot R_e}$) 
and are thus expected to give upper limits only to our global galaxy models. 

Increasing the timescale for SF we find that all models with ${\rm t_{\ast} 
\leq 5}$ Gyr and an average metallicity ${\rm \langle Z \rangle_V 
> \frac{1}{5} \cdot Z_{\odot}}$ reach the observed range of [MgFe]$_{\ast} 
\sim$ 0.4 - 0.6. Models with ${\rm t_{\ast} 
= 5}$ Gyr and ${\rm \langle Z \rangle_V 
\geq \frac{1}{2} \cdot Z_{\odot}}$ reach ${\rm Mg_2 \geq 0.30}$ and ${\rm Fe5335 
\geq 2.7}$ at ages $\geq 15$ Gyr. 

Revisiting our spiral-spiral merger models with interaction-triggered starbursts of 
various \hfill\break
strengths (Fritze -v. A. \& Gerhard 1995a,b) it turns out that in case of a strong 
burst (like the one in NGC 7252) 
merger remnants reach the 
observed [MgFe]$_{\ast} \sim$ 0.4 - 0.6 about 2 Gyr after the burst for 
${\rm \langle Z \rangle_V = Z_{\odot}}$ and about 4 Gyr after the burst for 
${\rm \langle Z \rangle_V = \frac{1}{2} \cdot Z_{\odot}}$. Merger 
remnants keep lower Mg$_2$ and Mgb than ellipticals at fixed M$_{\rm B}$ 
 \ for about 3 Gyr in agreement with observations 
(Schweizer \etal 1990, Bender \etal 1993) but can reach even the highest Mg$_2$ and 
Fe index values observed in ellipticals about 4 -- 5 Gyr after the burst.  

Spirals with SF truncation (via tidal stripping, sweeping or harassment) but 
without a burst -- which dynamically might transform into spheroidal galaxies -- 
reach the observed [MgFe]$_{\ast} \sim$ 
0.4 - 0.6, rapidly for ${\rm \langle Z \rangle_V = Z_{\odot}}$, i.e. early 
type spirals, but hardly for ${\rm \langle Z \rangle_V = \frac{1}{2} \cdot 
Z_{\odot}}$ (Sbc or later). They feature average elliptical galaxy values for Mg$_2$ 
and Fe5335 in case of early type spirals but values close to the lower limit of 
ellipticals (as appropriate for their low luminosity) in case of late type spirals 
(Fritze - v. A. 1997, {\sl in prep.}). 

All this is readily understood looking at the Lick calibrations (Gorgas \etal 1993). 
They directly show that [MgFe]$_{\ast} \gta 0.4$ for 
giants later than K0 (${\rm \frac{1}{2} \cdot Z_{\odot}}$) or later than 
G6 (Z$_{\odot}$), so that whenever these stars dominate the light 
[MgFe]$_{\ast}$ - values like those observed in giant Ellipticals are to 
be expected.
 
\section{Conclusions} 
In addition to giving a very brief overview on observations and current 
interpretations of abundances and abundance ratios in composite 
stellar populations I stressed that elliptical galaxies are really 
composite in terms of stellar metallicities (and ages) and require a chemically 
consistent description that takes into account the range of initial 
stellar abundances. I showed that while for stars or SSPs, like star clusters, 
absorption indices do reflect ISM abundance ratios at birth, this is not the case 
for CSPs. Conclusions concerning the timescale for SF or the IMF based on Mg$_2$ 
and Fe index data for elliptical galaxies or bulges may well be true, but need not.

% Begin acknowledgements
\acknowledgements{
I greatfully acknowledge a travel grant (Fr 916/4-1) from the Deutsche 
Forschungsgemeinschaft.}
% End acknowledgements

\end{document}